\documentclass{aa}

\usepackage{natbib}
\usepackage{psfig}
\usepackage{epsf}

\begin{document}

\title{Formation of massive skyrmion stars}

\titlerunning{Formation of massive skyrmion stars}

\author{S.B. Popov
            \inst{1,2}
             \and
 M.E. Prokhorov\inst{1}
         }

   \offprints{S. Popov}

\institute{Sternberg Astronomical Institute,
Universitetski pr. 13, 119992 Moscow, Russia\\
\email{polar@sai.msu.ru; mike@sai.msu.ru}            \and
   Universit\`a di Padova, Dipartimento di Fisica, 
via Marzolo 8, 35131, Padova, Italy\\
\email{popov@pd.infn.it}
}

   \date{}

\abstract{
As it is well known for stiff equations of state an existence of neutron 
stars with masses $\ga 2\, M_{\sun}$ is possible. Especially interesting 
possibility is opened if 
the equation of state based on the Skyrme theory is realized in nature. 
This equation of state was proposed recently by Ouyed and Butler.
We discuss different channels of formation of massive rapidly
rotating neutron stars. 
We use a population synthesis code to estimate numbers of massive neutron stars
on different evolutionary stages. 
A neutron star increases its mass by accretion from a secondary companion.
 Significant growth of a 
neutron star mass due to accretion is possible only for certain values of
initial parameters of the binary.
In this paper we show that significant part of massive neutron stars with
$M\ga 2\, M_{\sun}$ can be observed as millisecond radio pulsars,  as X-ray
sources in pair with white dwarfs, and as accreting neutron stars with very
low magnetic fields.  
\keywords{stars: neutron  -- stars: evolution -- stars:
statistics -- stars: binary -- X-ray: stars}}

\maketitle

\section{Introduction}

 Mass is one of the key parameter
for neutron star (NS) physics and astrophysics.
It can be measured with high precision in binary radio pulsar systems. 
Up to very recent time obtained results fell in a very narrow region
1.35-1.45~$M_{\sun}$ (\citealt{tc1999}). 
These values lie very close to the
Chandrasekhar limit for white dwarfs. 
Thus, for years $M=1.4\, M_{\sun}$ was considered to be
a standard value of a NS mass. Recently the range widened 
towards lower masses
thanks to the discovery of the double pulsar J0737-3039 
(\citealt{b2003}). 
One of the NSs in this system has $M=1.25\, M_{\sun}$ 
(\citealt{l2004}). 
Also the mass range expanded towards higher masses, though this result is
less certain.
There is only one NS in a binary radio pulsar system with  mass
significantly higher than the canonical value $1.4\, M_{\sun}$.
It is the pulsar J0751+1807  with the mass
$2.1^{+0.4}_{-0.5}$ (95\% confidence level) (\citealt{ns2004}).  
All others are consistent with the standard mass value on the 1-2-$\sigma$
level.
However,  small number of massive radio pulsars
can be a result of a selection effect(s).

 There are reasons to suspect an existence of significant number of
NSs with higher masses.
Evidence for such a proposal comes both from theory and observations.
Calculations of cooling curves of NSs suggest that some of these objects
might be more massive than known sources in radio pulsar systems
(see for example, \citealt{khy2001}) 
with $M$ up to 1.8~$M_{\sun}$ and probably more.
Modeling of supernova (SN) explosions also suggest the existence of NSs
with higher masses (\citealt{whw2002}).
Still models of NS thermal history and SN explosions do not requier masses
$M\ga 2 \, M_{\sun}$, but there are observational indications for their
existence.
 
Observationally high masses of NSs are mainly 
supported by data on X-ray binaries (we do not discuss here data based on
quasi-periodic oscillations measurements as they are model dependent).
Estimates for several systems give very high values: 
1.8--2.4~$M_{\sun}$ for Vela X-1\footnote{This range is based on the two
estimates given in (\citealt{q2003}): $1.88\pm 0.13$ and $2.27\pm 0.17 \,
M_{\sun}$.} 
(\citealt{q2003}), 
2.4$\pm 0.27$~$M_{\sun}$ for 4U 1700-37
(\citealt{c2002}; see also \citealt{h2003}; 
\citealt{vk2004}). Very recently \cite{s2004} 
presented observations of a low-mass X-ray binary 
2S~0921-630/V395~Car 
for which 1-$\sigma$ mass range for the compact object is 2--4.3~$M_{\sun}$.
Still it is necessary to note that 
such measurement are not as precise as the radio pulsar ones 
(for example,
at the 3-$\sigma$ level the mass of the NS in Vela X-1 is still compatible
with the standard value).

 The existence of NSs with 
$M\sim 2$--$2.4\, M_{\sun}$ 
is not in contradiction
with the present day theory of NS interiors. There are several models with
stiff equation of state (EOS) which allow an existence of NSs with masses
$\ga 2 \, M_{\sun}$ (see a review and references in \citealt{ha2003}). 
Here we will focus on so called {\it skyrmion stars}
(SkyS) as they are expected to be a kind of NSs with the highest value
of maximum mass ($M_\mathrm{max}$).

 In 1999 Ouyed and Butler discussed an EOS based on the model of 
\cite{s1962}.  A NS with such EOS has 
$M_\mathrm{max}$=2.95~$M_{\sun}$ 
even for a non-rotating configuration. 
Usualy maximum rotation can increase the limit by $\sim$~15--20\%.
Rapidly rotating SkyS were discussed by 
Ouyed (2002, 2004)\nocite{o2002}\nocite{o2004}, 
and for this  case $M_\mathrm{max}$=3.45~$M_{\sun}$ and
$R=23$~km (this model also has relatively large radii of NSs).
Such model is very interesting from the astrophysical point of view,
and it is important to discuss scenarios of formation of compact objects
with such high masses. Our goal in this note is to pick out evolutionary
tracks of binary systems which can lead to the formation of NSs with high
masses, and to discuss possible observational appearences of such systems and
their relative and absolute numbers in the Galaxy.
As we do not use explicitly any EOS in our calculations, then our results
can be applied to other stiff equation of state and even to low-mass black
holes (BHs).

In the next section we discuss evolutionary paths at the end of which a
massive NS can be formed. Then we give an estimate of the number of massive
NSs in the Galaxy. Finally we discuss our results and propose systems
which are more favorable to host massive NSs.

\section{Possible channels of massive neutron star formation}
\label{sec2}

As mass determination for NSs is possible only in binary systems%
\footnote{Note, that in principle there is a possibility to determine an 
isolated NS mass 
by microlensing effects, however, we do not touch this issue here.}
we focus on potentially observable stages of evolution of binary systems
in which a massive NS can form. Below we discuss possible ways of massive
NS formation. 

Since we are mostly interested in compact objects with rapid rotation 
(because  they can have higher maximum masses) 
it is necessary to follow evolution
in a binary as such objects cannot form from single stars
(\citealt{hw2003}),
so its necessary to study evolution of close binary systems. 
Except evolutionary tracks which lead to a formation of a massive NS in a
binary we follow  paths at the end of which an isolated massive NS can 
form. 
An appearence of a rapidly rotating single massive NS due to a binary 
evolution can be a result 
of a coalescence of two compact objects (NSs or white dwarfs --- WDs), or a
result of a more slowly  merging process in which a normal star is involved,
or a result of an evaporation of a low-mass secondary companion by an active
pulsar.
At some stages during its evolution a binary which finally is going to produce
an object of our interest  can be observed as an X-ray source,
that is why it is important to select evolutionary paths also for them.

The main output of a collapse of cores of massive stars are NSs with
$M\sim1.2$-1.5~$M_{\sun}$. This conclusion is supported both observationally
(\citealt{vk2004}) 
and theoretically (\citealt{tww1996}; 
\citealt{fk2001}; 
\citealt{whw2002}). 
Numerical models of collapse are not as precise as
necessary to determine the exact shape of a NS mass spectrum (for example
the amount of fallback is not well known), however, calculations show that
the formation of NSs with high masses is not favourable and most of them
should have $M\sim1.3$-$1.4\,M_{\sun}$. 

A discovery of a NS with $M\ga 1.8\,
M_{\sun}$ should mean that the mass was increased after formation of the
compact object during its evolution (if the mass is significantly higher
than 1.8~$M_{\sun}$ then such a conclusion seems to be inevitable).
Based on this proposition we call below as {\it massive} NSs with 
$M>1.8\, M_{\sun}$.      

A NS can increase its mass due to fallback, coalescence with another
NS, or accretion from a secondary companion. As we note above, the first way 
is not well studied, and we do not discuss it below. 
Oppositely coalescence of NSs is well understood (see \citealt{rrd2003} 
and references therein). 
The rate of NSs coalescence in the Galaxy is about
1 per $10^4$ yrs. As a result a rapidly rotating massive isolated NS 
(or a BH)
can form. This way of evolution also will not be discussed below. In the
following only binary evolution of a NS in pair with a normal star or a WD
will be studied.

At first for an illustration let us assume an isotropic collapse, ie. 
zero kick. Such an assumption is not realistic as most part of NS --
nearly all radio pulsars -- obtain at birth high additional velocity 
$\sim$100--1000~km~s$^{-1}$  
(\citealt{acc2002}). 
However it is much easier to understand main processes in a binary evolution
if one neglects kick. In addition, if a binary was not unbounded after a SN
explosion then an orbital eccentricity quickly decays after  the secondary
fills its Roche lobe. So, if at the moment we are not interested  in the
question of the binary survival then it is possible to neglect kick 
to simplify the explanation.   

 Let us start with a qualitative discussion (below in sec.~2.1 a more
detailed consideration is given). The most obvious channel to form a
rapidly rotating massive NS is an evolution in a low-mass or intermediate mass
binary (see, for example, recent calculations by \citealt{prp2002}). 
This path includes, for example, millisecond pulsars
(however it is not the only possible output). 

As we are interested here in
systems with high mass ratio (a massive primary produces a NS and the
secondary star has a low mass) it is necessary to consider three different
situations after the NS formation when the secondary fills its Roche lobe:
{\it i.)} a normal star can fill its Roche lobe without a common envelope
formation; {\it ii.)} a normal star can fill its Roche lobe with a common
envelope  formation; {\it iii.)} a WD fills its Roche lobe.

To fill the Roche lobe a normal secondary
star has to evolve further the main sequence stage. 
During its evolution prior to the Roche lobe overflow
the mass of the star is nearly constant
(see detailed tracks below). A common envelope is not formed if the normal
star is not significantly heavier than the NS. In this regime mass is not
lost from the binary system. For more massive secondaries formation of a
common envelope is inevitable, mass transfer is unstable. In this regime
significant fraction of the mass flow is lost from the system, so the mass
of the NS grows less effectively. 
It is only partly compensated by higher mass of
the donor.

After the common envelope stage an 
orbital separation becomes smaller, so later
on even a degenerated core of the secondary -- a WD -- can fill the Roche
lobe.

\subsection{Evolutionary tracks}

For our calculations we use the ``Scenario Machine'' code developed
at the Sternberg Astronomical Institute.
\footnote{Online materials are available at
http://xray.sai.msu.ru/sciwork/scenario.html and
http://xray.sai.msu.ru/~$\sim$mystery/articles/review/.}
Description of most of parameters of the code can be found
in (\citealt{lpp1996}). 
Below we mention those which are the most important for us here:

\begin{itemize}
\item All NSs are born with $M=1.4\, M_{\sun}$.
\item At the common envelope stage a hypercritical accretion
(with $\dot M$ much larger than the Eddington value) is possible.
\item During accretion the magnetic field of a NS decays down to the value
which cannot prevent rapid (maximum) rotation of the NS.
\item Oppenheimer-Volkoff mass 
of a rapidly rotating NS
(the critical mass of a BH formation) 
is assumed to be 3.45~$M_{\sun}$ according to \cite{o2004}. 
\end{itemize}

For zero kicks we distinguish two groups of tracks which produce
massive NSs. A typical track from the first group has initial value of the
semimajor axis $a=290\, R_{\sun}$ and star masses $M_1=10.5\, M_{\sun}$,
$M_2=2\,M_{\sun}$ (fig.~1 left)%
\footnote{
Colored version of the figure in high resolution is avalable on the Web:
http://xray.sai.msu.ru/$\sim$polar/html/publications/ouyed/
}.
After the massive component leaves the main
sequence it expands and fills its Roche lobe. As a result the common
envelope stage sets on. During this stage the orbit shrinks by more than an
order of magnitude, and the primary looses about 3/4 of its mass and becomes
a low-mass helium SN progenitor. After the SN explosion the orbit
has low eccentricity and $a\sim 7$--8~$R_{\sun}$. Mass of the secondary 
is not changed during these stages of the evolution.

 Till the secondary fills its Roche lobe the NS is at the stages of {\it
ejector} and {\it propeller} (see for example \citealt{l1992}
for stages descriptions). 
During these stages the magnetic field is assumed
to be constant. Stage durations can be found in 
(\citealt{lpp1996})\footnote{The subsonic propeller stage 
is not taken into account as for
binaries with big accretion rates it is very short.}. 

After the secondary fills the Roche lobe the NS
starts to accrete. At that moment the mass ratio is about 0.7 
(the NS is lighter) and a mass transfer
is stable with nearly zero mass loss from the system. Up to equalizing
of components masses matter transfer goes on a thermal time scale, after
equalizing -- on a nuclear. Process of accretion can be stopped because of 
a switching on of a millisecond radio pulsar. It happens when the donor's
mass is $\sim 0.1\, M_{\sun}$. The remnant of the secondary companion then can
be evaporated completely, while the evaporation is proceeding
the systems looks like the famous ``Black widow'' pulsar 1957+20
(and its twin PSR J2051-0827). 
If accretion is not stopped then
it continues till a planet-like (Jupiter mass) companion remains. 
As we see the final stage of such an evolution is a ``single'' massive rapidly
rotating NS. In both cases the final mass of a NS can reach
3.2--3.3~$M_{\sun}$. We can observe such a system at the stage of accretion
which lasts 90\% of the evolution. Masses of NSs in these accreting systems
can be in the range from the initial mass (1.4~$M_{\sun}$ in our case) up to
3.2--3.3~$M_{\sun}$. Orbits can be relatively wide.

 The described evolutionary channel appears to be narrow in a sense that
small changes in the initial conditions do not allow a massive NS formation.
Also uncertain parameters of the common envelope stage can significantly
influence this path.

Ranges of 
initial parameters of evolutionary tracks from the second group are given in 
the table 1. 
We give maximal and minimal values for two types of tracks (2a and 2b)
which differ by the final stages of evolution.
The orbital period, $P_\mathrm{orb}$, is given in the table 1 just for
an illustration. In our calculations we use masses and semimajor axes.
So the values of $P_\mathrm{orb}$ given in the table are simply calculated
using maximum masses and minimum semimajor axes for the shortest period, 
and, oppositely, minimum masses and maximum axes for the longest period.
Due to that ranges for $P_\mathrm{orb}$ for tracks 2a and 2b intersect.

A typical representative of the 2a subgroup has the following 
initial parameters:
$a=300\, R_{\sun}$, $M_1=12\, M_{\sun}$, $M_2=4\, M_{\sun}$. 
The main difference form the first group of tracks is a more massive
secondary companion. Because of that the common envelope during the first
mass transfer is less effective, and after a SN a system with $a=170\,
R_{\sun}$ and low eccentricity is formed (the mass of the secondary is not
changed). Later the secondary fills the Roche lobe. Mass ratio is high, mass
transfer is unstable and the common envelope forms. At the end of the common
envelope stage the secondary becomes a WD with $M\sim0.8\, M_{\sun}$, and the
orbital separation diminishes down to $5\, R_{\sun}$. During the common
envelope stage the NS increases its mass up to $\sim 2.3\, M_{\sun}$ (for more
massive donors mass loss from the system is more effective, so in such cases
the NS mass can be lower: $\sim 1.9\, M_{\sun}$).

After the formation of a binary consisting of a NS and a WD 
the evolution
in the second group can  take one of two different paths.
For some tracks (2a) from the second group the time of  rapprochement of
the components due to gravitational wave emission
is too long, so  there is no Roche lobe overflow. 
Systems with smaller orbital separation have enough time to approach
to each other close enough for the beginning of WD overflow.  
This situation corresponds to the initial parameters $a=200\, R_{\sun}$,
$M_1=12\, M_{\sun}$, $M_2=4\, M_{\sun}$ (track 2b in the table 1).

The main
difference between tracks 2a and 2b is smaller orbital separation in the
latter case. Track 2b is similar to the one on the right panel of fig.1, but
after the common envelope semiaxis of the system is just $\sim3\, R_{\sun}$.
A WD has enough time to fill the Roche lobe and completely transfer its mass
to the NS.   
At the end we have a single rapidly rotating NS. The NS mass
for this case is increased up to $\sim 3\, M_{\sun}$.
Stages with a WD are shown in the box as they distinguish the track 2b
from 2a.

 When a semimajor axis is larger than $a\sim 670\, R_{\sun}$ 
the second common envelope results
in NS--star merging, so the Thorne-Zytkow object is formed. Its evolutionary 
path is not very clear. A formation of a massive NS and a formation of a BH
are both possible. We do not include this possibility into our calculation.

\begin{table}
\caption{Parameters for tracks from the second group}
\begin{tabular}{crrl}
\hline\hline
parameter & $\min$ & $\max$ & width \\
\hline
\multicolumn{4}{l}{Track 2a} \\
$a$ & 279$R_{\sun}$ & 670$R_{\sun}$ & 0.20 \\
$M_1$ & 10.3$M_{\sun}$ & 12.8$M_{\sun}$ & 0.054 \\
$M_2$ &  3.9$M_{\sun}$ &  6.7$M_{\sun}$ & 0.13 \\
$P_\mathrm{orb}~{}^{(*)}$ & $123^\mathrm{d}$ & $537^\mathrm{d}$ & \\
&&&\\
\multicolumn{4}{l}{Track 2b} \\
$a$ & 135$R_{\sun}$ & 279$R_{\sun}$ & 0.17 \\
$M_1$ & 10.3$M_{\sun}$ & 12.4$M_{\sun}$ & 0.046 \\
$M_2$ &  3.9$M_{\sun}$ &  7.4$M_{\sun}$ & 0.15 \\
$P_\mathrm{orb}~{}^{(*)}$ & $41^\mathrm{d}$ & $144^\mathrm{d}$ & \\
\hline
\multicolumn{4}{p{6cm}}{$^{(*)}$ $P_\mathrm{orb}$ is given just as an illustration, see the
text.}
\end{tabular}
\end{table}

\begin{figure}
\hbox to\columnwidth{
\hfill{\epsfysize=12cm\epsfbox{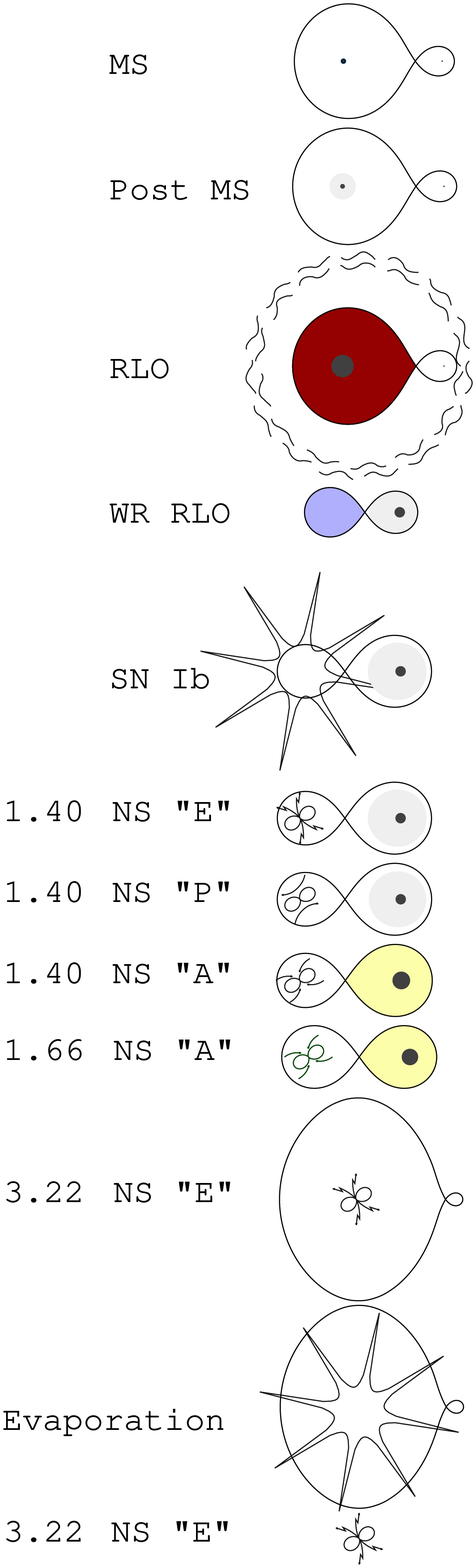}}%
\hfill\hfill%
{\epsfysize=12cm\epsfbox{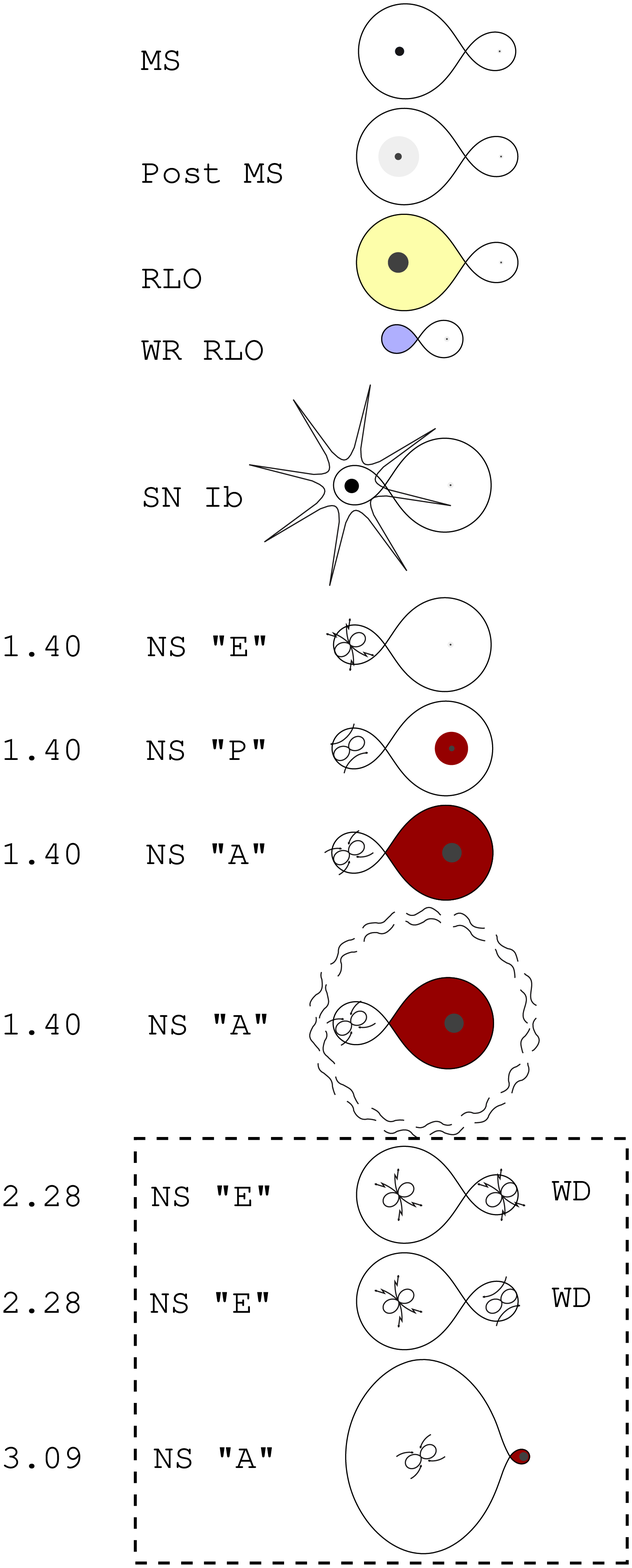}}\hfill%
}
\caption{Evolutionary tracks for massive NS formation. In the left panel we
show a typical track from the first group. The first mass transfer (from the
primary) results in a common envelope formation due to high mass ratio.
Accretion onto a NS from the secondary companion proceeds stably without a
common envelope. In the right panel we show an evolutionary path of a system
from the second group. This track differs by a higher mass of a secondary
companion. Because of this difference the first mass transfer goes on
without a common envelope. A NS gathers an additional mass during one or two
episodes of accretion. If the orbital separation is not very large ($\sim
200\, R_{\sun}$, see text) then 
at first the NS accretes from a normal secondary filling its Roche
lobe, and then from a WD (this stage is shown in the dashed frame). 
For wider systems
the evolution stops after the mass transfer from the normal secondary star
(ie. before the frame).     
On the left from each track we indicate evolutionary stage (in notation from
Lipunov, Postnov \& Prokhorov 1996) and NS masses. 
}
\end{figure}

\subsection{Evolutionary tracks with kicks}

Above we discuss two families of tracks with zero kicks which result in
massive NSs formation. However, it is necessary to include kicks as they are
a general property of a NS formation. A kick can change orbital parameters
after a SN explosion, it can even make the system unbounded.    If after a
SN (and after a brief period of circularization of an orbit)
we obtain in our calculations a system with parameters in the range
which was obtained above for the zero kick, then the following history of the
system should be the same as described in sec.~2.1.  

An additional velocity which a NS obtains at birth can change the range
of initial parameters that are necessary for a massive NS formation.
Especially it is important to estimate if ranges for $M_1$, $M_2$ and $a$
are changed significantly or not.
As a kick velocity and a NS mass in our calculations 
are assumed to be independent on a mass
of an exploding star (see below sec. 4.1)
we do not expect that a range of  masses of primaries
should be modified. The same can be said about the range of initial masses
of secondaries because these stars do not suffer any important evolutionary 
changes before a SN expolsion.
Since a kick can dramatically change the orbital parameters the 
situation is different for the initial orbital separation range.  
For example, with a kick systems wider then the ones discussed in sec.~2.1
can still form massive NSs.

In the next section we present results of our calculations of population
synthesis of massive NSs for both scenarios.

\section{Estimate of observable number of massive neutron stars in the
Galaxy}

\begin{table}
\caption{Fractions of massive NSs at different stages}
\begin{tabular}{lrr}
\hline\hline
Stage & with kick & without kick \\
\hline
Ejector                & 0.32 & 0.39 \\
Propeller + Georotator & 0.02 & 0.08 \\
Accretor               & 0.66 & 0.53 \\
Hypercritical stages   & $5\cdot10^{-6}$ & 0 \\
\hline  
\end{tabular} 
\end{table}

To estimate the number of massive NSs in the Milky Way we run several sets
of population synthesis calculations for the ranges of initial parameters
which correspond to the two groups of tracks described above. Each run 
includes calculations of $10^6$ individual binary evolutionary tracks.

We run the model for zero kick velocities and for non-zero ones. For the
latter case we use the distribution similar to the one suggested in
(\citealt{acc2002}). 
We use bimodal distribution with equal fraction of objects
in each mode. An average velocity in the first mode is 175 km~s~$^{-1}$
and in the second it is  750 km~s~$^{-1}$, distribution in each mode is
maxwellian. 

For the scenario without kick we proceed as follows.
For the second group of tracks we used ranges indicated in the table 1.
Width given in the table is calculated as 0.5(max-min)/(max+min).
For the first family of tracks we used the range for $a$ from 230 to
346~$R_{\sun}$, 
for $M_1$ from 8.4 to 12.6~$M_{\sun}$, and for $M_2$ from 1.6 to
2.4~$M_{\sun}$.

For the scenario which takes into account 
an additional velocity gained by a NS
at birth we used wider range of initial semimajor axis: from 200 to 2000
$R_{\sun}$.   Masses are chosen in the same way as for the zero kick variant.

The results of the calculations for non-zero kick are the following
(we assume the total number of all NSs in the Galaxy as $10^9$, and the
galactic age as $1.5\, 10^{10}$~yrs).
In the first channel (fig.~1 left panel) we do not obtain significant number
of massive NSs. Most of these objects are formed in the second channel.  
Formation rate of massive NSs was found to be $6.7 \, 10^{-7}$~yrs$^{-1}$.
This corresponds to $\sim10\,000$ of these compact stars in the Galaxy.  
For zero kick the formation rate is larger $4 \, 10^{-6}$~yrs$^{-1}$,
so the total number is $\sim 60\, 000$. 

Certainly only a fraction of massive NSs at any given moment passes through
stages which are  observable, ie. the {\it accretor} stage and the
stage of radio pulsar. 
Some of these objects are at stages of
{\it ejector} and {\it propeller or georotator}. All three of them
are not favourable for detection%
\footnote{We note, that the {\it ejector} stage does not coinside with the
radio pulsar stage, but includes it as a substage. 
So here we are speaking about
non-detectability of {\it ejectors} which are not active as radio pulsars.
See for example (\citealt{l1992}) or (\citealt{lpp1996}) for more details.}.
In the table 2 we give fractions of massive NSs on each stage. 
It is clear that {\it accretors} are more numerous (but the number of
massive NSs at the stage of superEddington accretion is negligible). 

For the non-zero kick model
about 25\% of accreting massive NSs have normal stars
as secondaries, the rest 75\% have WD companions. For zero kick nearly all
massive NSs accrete from WDs which fill their Roche lobes.

Mass distributions for both scenarios are shown in the fig. 2. 
Note, that all small details in the figure are due to statistical noise
(for example, the first peak on the rising part of the dashed curve, or the
middle peak on the solid one). The only important details are the two peaks
at $M\sim 2.3 \, M_{\sun}$ and $M\sim 3.1 \, M_{\sun}$,
which correspond to tracks 2a and 2b (see the right panel of Fig.~1 and
table~1).  
As we found only two groups of tracks which lead to the 
formation of massive NS
only results obtained for these groups are shown. All contributions from
other types of tracks are not considered here and in fig.~3.   

Finally, in the last
figure we represent luminosity distributions. 
For the scenario with non-zero kick about 1/2 of massive NSs have $M>
2.5\, M_{\sun}$. Taking all together we can conclude that in the Galaxy
there are several
thousand of accreting massive NSs with luminosities 
$10^{34}\la L\la 10^{36}$~erg~s$^{-1}$.

\begin{figure}
\centerline{\psfig{figure=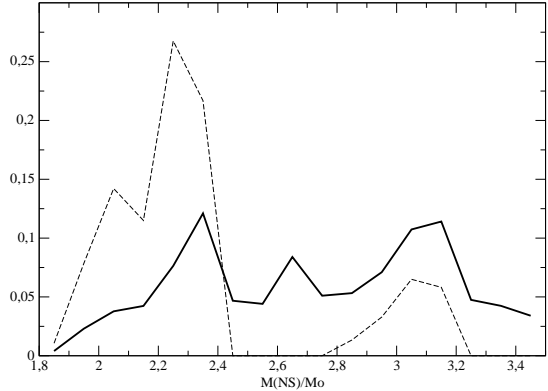,width=0.99\hsize,angle=270}}
\caption[]{Mass distribution of NSs. As we are interested only in the massive
population  we do not show the results for compact
objects with $M<1.8\, M_{\sun}$. Upper mass limit corresponds to SkyS
 with maximum rotation (Ouyed 2004). 
The dashed line represents results for the scenario with zero
kick. The solid line -- non-zero kick. Left peaks for both distributions
correspond to NSs with a single episode of accretion. Right peaks are formed
by NSs which also increased their masses via accretion from WDs. 
Distribution were normalized to unity, ie. an area below each line is equal
to one.
}
\label{hyst}
\end{figure}

\begin{figure}
\vbox{\psfig{figure=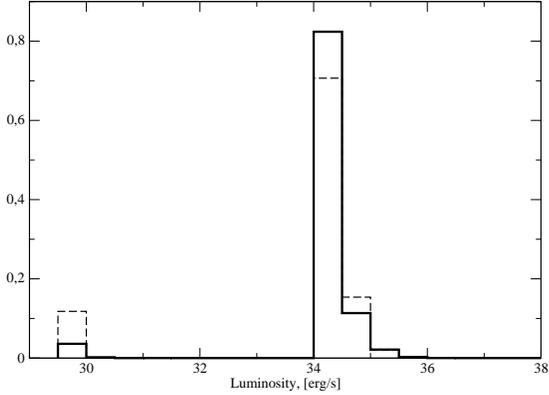,angle=270,width=\hsize}}
\caption[]{Luminosity distribution of accreting massive NSs. The left bin
includes all sources with $L<10^{30}$~erg~s$^{-1}$. The dashed line
corresponds to the scenario with zero kick. The solid line -- non-zero kick.
In the ranges $10^{30}\la L \la 10^{34}$ erg~s$^{-1}$
and  $10^{36}\la L \la 10^{37}$  erg~s$^{-1}$
the number of systems is not equal to zero, however, it is very small. All
distributions are normalized to unity.}
\label{hyst}   
\end{figure}

\section{Discussion and additional comments}

 Here at first we notice some uncertainties of the scenario. 
Then we briefly discuss a possibility of massive NS formation in globular
clusters, low-mass BHs, and types of sources which can host massive NSs.

\subsection{Correlations between initial parameters of neutron stars}

 The scenario of binary evolution that we use has different types of
uncertainties. Here we touch just one of them -- possible correlations
between parameters of the scenario.

 In our calculations we 
assumed that such initial parameters of NSs as spin period, magnetic
field, mass, and velocity are uncorrelated with each other. 
The reason for this
assumption is trivial: there is no direct indication on such correlations.
However, theorists suggested plethora of them. We just give a list
(probably incomplete) of
possibilities and corresponding references to  original papers.

\begin{itemize}
\item Spin -- magnetic field (\citealt{td1993}).
\item Magnetic field -- mass (\citealt{petal02}; \citealt{hws04}).
\item Internal structure -- velocity (\citealt{bp04}). 
\item Binarity -- velocity (\citealt{petal2004}).
\item Core mass -- velocity (\citealt{setal2004}).
\end{itemize} 

 As here we mainly speak about masses and kick velocities, let us 
make a short comment on last two items.  Our calculations may not be strongly
influenced by such  correlations. The reason is that the mass added during
accretion is much larger then a difference in initial masses. I.e., as we
need to accrete nearly two solar masses to obtain the most massive SkyS 
at maximum rotation, we can safely forget about the range of initial masses.
That's why in our calculations we even assume, that all NSs have the same
initial mass.  For the same reason we can neglect the second item in the
list above. 

 As here we deal with systems with high accretion rates, and the magnetic
field is assumed to decay due to accretion, the first item also is not
very important for us. Even if a NS was a magnetar in its early years, later
on after field decay it can follow a normal evolutionary path.

 We do not discuss here a possibility of phase transition in a NS interiors.
Formation of a quark phase due to mass increase can significantly influence
the following history of a binary system, but we just neglect it here as in
that case formation of massive NSs is impossible. 
 
\subsection{Globular clusters}

 All evolutionary tracks that we present above correspond to binary
evolution in the Galaxy, and so they cannot be directly applied to globular
clusters. However, we are mainly interested in systems which 
after a formation of a NS appear to be stiff, ie. orbital velocities in
binaries are larger than a velocity dispersion in a cluster. It is also true
during the following evolution of a system, and it can be violated only at
the stage of Roche lobe overflow by a WD. 
We can conclude that a dynamical
influence of the globular cluster stellar population should not destroy
systems under discussion. Still duration of various evolutionary stages can
be different in clusters and in the galactic disc, 
and so our estimates of relative fractions cannot be valid
for globular clusters.     

It is possible to speculate that as the formation rate of millisecond
pulsars is enhanced in globular clusters then the formation rate
of massive NSs can also be higher there than in the disc of the Galaxy.
It is an important question because massive NSs from globular clusters can
enrich the disc population of these objects (so, our calculations which
neglect this contribution give a lower limit on the number of massive NSs
in the Galaxy). 
In our opinion millisecond radio pulsars and X-ray sources in globular
clusters can be good candidates for a search of massive NSs.

\subsection{Low-mass black holes}

As it was described in the Introduction at present all well determined
values of NS masses lie below 
$\sim$1.5~$M_{\sun}$, on the other hand most of BH
mass determinations lie around values 6-10~$M_{\sun}$ 
(\citealt{z2004}). So there
is an indication on a gap in intermediate mass range.    
Briefly we can say, that accretion cannot fill this gap if, as it is
standardly assumed, NSs are formed with $M\la 2 \, M_{\sun}$ and BHs
are formed with $M\ga 5 \, M_{\sun}$.

If an EOS of NSs with a very high $M_\mathrm{max}$ is realized in nature,
then up to $\sim 3\, M_{\sun}$ or even further, in the case of maximum
rotation, we can find NSs. Otherwise the gap above $\sim 2 M_{\sun}$
should be populated only by BHs. Even in the case of the EOS discussed by
\cite{ob1999} low-mass BHs can form from rapidly rotating massive NSs as
they slow down. 

Fig. 2 (the solid line) clearly shows that the number of low-mass BHs (or
any other type of compact objects) with 
$M \ga 3.2 $ in our scenarion is small. However, if $M_\mathrm{max}$ is
$\sim 2 \, M_{\sun}$, and if a binary is not significantly influenced during
a BH formation (i.e. accretion continues), 
then the number of BHs with $M_\mathrm{max}\la M
\la 3.2 \, M_{\sun}$ is non-negligible.\footnote{Mass growth of NSs and BHs
in close binaries is also discussed in (\citealt{betal2005}) and 
in (Bogomazov et al., in preparation).} 

There are several examples of binary systems with an estimate of a  
mass of a compact object $\sim3$-$4\,M_{\sun}$ 
(\citealt{or2004};  \citealt{s2004}).
These objects are considered as BH candidates.
In principle such objects can be formed in the frame of the scenario
discussed above after a mass of a NS exceeds the Oppenheimer-Volkoff limit.

\subsection{Possible candidates}

 Main astrophysical manifestations of massive NSs  are the same as
for normal NSs: X-ray sources and radio pulsars. However, there are
differences. Very massive NSs should have short spin periods as they get an
additional mass by accretion which spin-up them and provoke magnetic field
decay.
\footnote{If a NS has a very short spin period then pulsations in an X-ray
source can be undetectable as it is observed for many low-mass X-ray binaries.}  
Of course a given millisecond pulsar can contain a NS with a normal mass.
Presence of a low-mass degenerated companion (a WD)
can be an indication that the system can hide a massive NS. 
One can mention another  additional signature of a massive NS -- very low
magnetic field. 

If the magnetic field is very small, then the Alfven radius becomes less
than the NS radius, and the accretion disk can nearly approach the NS surface.
This situation takes place when
$$
        B \la 2\cdot10^{9}~\mathrm{G}\, 
        \dot M_{-8}^{1/2} m^{1/4} r_{10}^{-5/4}\,,
$$
here $\dot M_{-8}\equiv \dot M/10^{-8}$~M$_\odot$/yr, $r_{10}\equiv
r_{\mathrm{NS}}/10$~km and $m$ --- mass of NS in Solar units. Thus,
the magnetic field strenght can be $\la 10^{9}$~G for Eddington accretion rate.
In that case a formation of a
boundary layer is favorable, and in the NS spectrum an additional thermal
component can be present (\citealt{is1999}). 
 For massive NSs including Ouyed EOS radius of a star is smaller
than the distance to the last stable orbit, so the disc cannot
actually smoothly approach the surface but qualitative properties
of a spectrum will remain the same.

All these consideration can be summarized in a list of types of objects
which can contain massive NSs.

\begin{itemize}
\item X-ray sources with weak pulsations with signatures of a boundary
layer;
\item millisecond X-ray pulsars with WD companions;
\item millisecond radio pulsars with WD companions;
\item other kinds of millisecond X-ray pulsars;
\item other kinds of millisecond radio pulsars.
\end{itemize}
By "other kinds" we mean millisecond pulsars with other types of companions
or isolated (but old) ones. Note, that we do not include into our
calculations secondary companions with very low initial mass (brown dwarfs).
However, such systems  cannot produce massive rapidly rotating
SkyS as the total amount of accreted matter is not sufficient.
NSs with very low-mass companions  like
the millisecond accreting pulsar SAX J1808.4-3658 or like ``black
widow''-like radio pulsars can be produced in our scenario via evaporating
degenerated or non-degenerated secondaries (see discussions 
on the evolution of this source in \citealt{ea1999};
\citealt{bc2001} and references therein). 

Unfortunately our calculations cannot provide exact numbers of objects of
each type. Uncertainties are connected with influence of population of
sources from globular clusters and with uncertainties of the scenario
itself. For example we absolutely do not take into account influence of
rotation on the evolution of normal stars (see \citealt{l2003}).

\cite{o2002, o2004} discussed three binary systems as possible candidates
to massive SkyS: 4U 0614+09, 4U 1636-53, 4U 1820-30. From the point of view
of evolutionary scenarios discussed above all three really can contain a
massive NS.  4U 1820-30 is especially interesting. The orbital period of the
system is only 11 minutes which means that the secondary is a low-mass
helium star (see \citealt{bs2004} and
references therein). However, this sources is situated in a globular cluster,
and so our considerations should be applied with care.

\section{Conclusions}

We discussed possible channels of massive NS formation. If the EOS based on
the Skyrme model 
suggested by \cite{ob1999} is realized in
nature then these objects can be SkyS with masses up to 3.45~$M_{\sun}$
for maximum rotation. The estimated numbers of these sources in the Galaxy
is high enough. Most favourable candidates are X-ray binaries with WDs as
donors, millisecond radio pulsars in pair with WDs and accreting NSs with
very low estimated magnetic field.
If no of so massive NSs are found in these systems
then the SkyS EOS has to be rejected. 

\begin{acknowledgements}

We thank drs. I. Bombaci, R. Ouyed, and the unknown referee for comments 
on the text of the manuscript.
SP thanks prof. J. Zdunik for discussions on the EOS.
This work was supported by the RFBR grants 04-02-16720 and 03-02-16068.

\end{acknowledgements}


\begin{thebibliography}{}


\bibitem[\protect\citeauthoryear{Arzoumanian et al.}{2002}]{acc2002}
Arzoumanian, Z., Chernoff, D.F., \& Cordes, J.M. 2002, ApJ, 568, 289

\bibitem[\protect\citeauthoryear{Ballantyne \&  Strohmayer}{2004}]{bs2004}
Ballantyne, D.R., \& Strohmayer, T.E. 2004, ApJ, 602, L105
  
\bibitem[\protect\citeauthoryear{Bildsten \& Chakrabarty}{2001}]{bc2001}
Bildsten, L., \& Chakrabarty, D. 2001,
ApJ, 557, 292

\bibitem[\protect\citeauthoryear{Bogomazov et al.}{2005}]{betal2005}
Bogomazov, A.I., Abubekerov, M.K., Lipunov, V.M., \& Cherepashchuk, A.M.
2005, Astronomy Reports 49 (in press)

\bibitem[\protect\citeauthoryear{Bombaci \& Popov}{2004}]{bp04}
Bombaci, I., \& Popov, S.B. 2004, A\&A, 424,  627

\bibitem[\protect\citeauthoryear{Burgay et al.}{2003}]{b2003} 
Burgay, M., D'Amico, N., Possenti, A. et al. 2003, Nature,  426, 531

\bibitem[\protect\citeauthoryear{Clark et al.}{2002}]{c2002}
Clark, J.S., Goodwin, S.P., Crowther, P.A., et al. 2002,
A\&A, 392, 909

\bibitem[\protect\citeauthoryear{Ergma \& Antipova}{1999}]{ea1999}
Ergma, E., \& Antipova, J. 1999, A\&A, 343, L45

\bibitem[\protect\citeauthoryear{Fryer \& Kalogera}{2001}]{fk2001}
Fryer, C.L., \& Kalogera, V. 2001,
ApJ, 554, 548 

\bibitem[\protect\citeauthoryear{Haensel}{2003}]{ha2003}
Haensel, P. 2003, in: ``Final Stages of Stellar Evolution'',
ed. C. Motch, \&  J.-M. Hameury, EAS Publications Series 7, 249

\bibitem[\protect\citeauthoryear{Heger et al.}{2003}]{hw2003}
Heger, A.,  Woosley, S.E.,  Langer, N., \&  Spruit, H.C. 2003,
in: Proc. of IAU Symp. 215 ``Stellar evolution'', 
ed. A. Maeder, P. Eenes, ASP, San Francisco
 (in press) [astro-ph/0301374]

\bibitem[\protect\citeauthoryear{Heger et al.}{2004}]{hws04}
Heger, A., Woosley, S.E., \& Spruit, H. 2004, ApJ, submitted
[astro-ph/0409422]

\bibitem[\protect\citeauthoryear{Heineke et al.}{2003}]{h2003}
Heinke, C.O., Grindlay, J.E., Lloyd, D. A., \& 
Edmonds, P. D. 2003, ApJ 588, 452

\bibitem[\protect\citeauthoryear{Hobbs et al.}{2004}]{h2004}
Hobbs, G.,  Faulkner, A., Stairs, I. H. et al. 2004, 
MNRAS, 352, 1439

\bibitem[\protect\citeauthoryear{Inogamov \& Sunyaev}{1999}]{is1999}
Inogamov, N.A., \& Sunyaev, R.A. 1999, Astronomy Letters, 25, 269

\bibitem[\protect\citeauthoryear{Langer et al.}{2003}]{l2003}
Langer, N., Yoon, S.-C., Petrovic, J., \& Heger, A. 2003,
in: Proc. of IAU Symp. 215 ``Stellar Rotation'',
ed. A. Maeder, P. Eenes, ASP, San Francisco (in press)
[astro-ph/0302232]

\bibitem[\protect\citeauthoryear{Lipunov}{1992}]{l1992}
Lipunov, V.M. 1992,
``Astrophysics of neutron stars'', Springer-Verlag, Berlin

\bibitem[\protect\citeauthoryear{Lipunov et al.}{1996}]{lpp1996}
Lipunov, V.M. Postnov, K.A., \& Prokhorov, M.E. 1996,
Astrophys. and Space Science Rev.,  9, 1

\bibitem[\protect\citeauthoryear{Lyne et al.}{2004}]{l2004}
Lyne, A., Burgay, M., Kramer, M. et al. 2004, Science 303, 1153

\bibitem[\protect\citeauthoryear{Nice \& Splaver}{2004}]{ns2004}
Nice, D.J., \& Splaver, E.M. 2004, astro-ph/0411207


\bibitem[\protect\citeauthoryear{Orosz et al.}{2004}]{or2004}
Orosz,  J.A.,  McClintock, J.E., Remillard, R.A., \& Corbel, S. 2004,
 astro-ph/0404343 

\bibitem[\protect\citeauthoryear{Ouyed}{2002}]{o2002}
Ouyed, R. 2002, A\&A,  382, 939  

\bibitem[\protect\citeauthoryear{Ouyed}{2004}]{o2004}
Ouyed, R. 2004, astro-ph/0402122 

\bibitem[\protect\citeauthoryear{Ouyed \& Butler}{1999}]{ob1999}
Ouyed, R., \& Butler, M. 1999,
ApJ, 522, 453 

\bibitem[\protect\citeauthoryear{Podsiadlowski et al.}{2002}]
{prp2002}
Podsiadlowski, Ph., Rappaport, S., \&  Pfahl, E.D. 2002,
ApJ, 565, 1107

\bibitem[\protect\citeauthoryear{Podsiadlowski et al.}{2004}]
{petal2004}
Podsiadlowski, Ph.,  Langer, N., Poelarends, A.J.T., et al. 2004, ApJ 612, 1044

\bibitem[\protect\citeauthoryear{Popov et al.}{2002}]{petal02}
Popov, S.B., Prokhorov, M.E., Colpi, M., Treves, A., \& Turolla, R. 
 2002, in Proc. of the Third International
Sakharov Conference on Physics, ed. A. Semikhatov et al. , Scientific
World: Moscow, p 420
[astro-ph/0210688]

\bibitem[\protect\citeauthoryear{Quaintrell et al.}{2003}]{q2003}
Quaintrell, H., Norton, A. J., Ash, T. D. C. et al. 2003, A\&A,  401, 313

\bibitem[\protect\citeauthoryear{Rosswog et al.}{2003}]
{rrd2003}
Rosswog, S.,  Ramirez-Ruiz, E., \&  Davies, M.B. 2003,
MNRAS, 345, 1077 

\bibitem[\protect\citeauthoryear{Shahbaz et al.}{2004}]{s2004}
Shahbaz, T.,  Casares J., Watson, C. et al. 2004, astro-ph/0409752

\bibitem[\protect\citeauthoryear{Shakura \& Sunyaev}{1987}]{ss1987}
Shakura, N.I., \& Sunyaev, R.A., 1987, Adv. In Space Res. 8, 135

\bibitem[\protect\citeauthoryear{Scheck et al.}{2004}]{setal2004}
Scheck, L., Plewa, T., Janka, H.-Th., Kifonidis, K., \& M\"uller, E. 2004,
PRL 92, 1103

\bibitem[\protect\citeauthoryear{Skyrme}{1962}]{s1962}
Skyrme, T.H.R. 1962, Proc. R. Soc. London A,  260, 127

\bibitem[\protect\citeauthoryear{Thompson \& Duncan}{1993}]{td1993}
Thompson, C., \& Duncan, R.C. 1993, ApJ 408, 194

\bibitem[\protect\citeauthoryear{Thorsett \& Chakrabarty}{1999}]{tc1999}
Thorsett, S.E., \& Chakrabarty, D. 1999, ApJ, 512, 288

\bibitem[\protect\citeauthoryear{Timmes et al.}{1996}]{tww1996}
Timmes, F.X., Woosley, S.E., \& Weaver, T.A. 1996,
ApJ, 457, 834 

\bibitem[\protect\citeauthoryear{van Kerkwijk}{2004}]{vk2004}
van Kerkwijk, M.H. 2004, astro-ph/0403489

\bibitem[\protect\citeauthoryear{Woosley et al.}{2002}]{whw2002}
Woosley, S.E., Heger, A., \& Weaver, T.A. 2002,
Rev. Mod. Phys.,  74, 1015 

\bibitem[\protect\citeauthoryear{Kaminker et al.}{2001}]{khy2001}
Kaminker, A.D., Haensel, P., \&  Yakovlev, D.G. 2001,   
A\&A, 373, L17

\bibitem[\protect\citeauthoryear{Ziolkowski}{2004}]{z2004}
Ziolkowski, J. 2004, Chinese Journal of Astronomy and Astrophysics (in
press) [astro-ph/0404052]

\end{thebibliography}
\end{document}